\documentclass[apjl]{emulateapj}
\usepackage{natbib}

\shorttitle{TP-AGB calibration}
\shortauthors{NOEL ET AL.}

\begin{document}


\title{Calibrating stellar population models with Magellanic Cloud star clusters}

%

\author{N. E. D. No\"{e}l\altaffilmark{1,}\altaffilmark{2}, L. Greggio\altaffilmark{3}, A. Renzini\altaffilmark{3}, C. M. Carollo\altaffilmark{1} and C. Maraston\altaffilmark{4}
}

\altaffiltext{1}{ETH Z\"{u}rich, Institute for Astronomy, Wolfgang-Pauli-Strasse 27, Building HIT, Floor J, CH-8093 Zurich, Switzerland}
\altaffiltext{2}{Department of Physics, University of Surrey, Guildford, GU2 7XH, UK}
\altaffiltext{3}{INAF-Osservatorio Astronomico di Padova, Vicolo dell'Osservatorio 5, I-35122 Padova, Italy}
\altaffiltext{4}{University of Portsmouth, Dennis Sciama Building, Burnaby Road, Portsmouth PO1 3FX, UK}




\begin{abstract}

Stellar population models are commonly calculated using star clusters as calibrators for those evolutionary stages that depend on free parameters. 
However, discrepancies exist among different models, even if similar sets of calibration clusters are used. With the aim of 
understanding these discrepancies, and of improving the calibration procedure, we consider a set of 43 Magellanic Cloud (MC) clusters taking age and photometric information from the literature. We carefully assign ages to each cluster based on up-to-date determinations ensuring that these are as homogeneous as possible. To cope with statistical fluctuations, we stack the clusters in five age bins deriving for each of them integrated luminosities and colors.  We find that clusters become abruptly red in optical and optical-IR colors as they age from $\sim 0.6$ to $\sim$1 Gyr, which we interpret as due to the development of a well-populated thermally pulsing asymptotic giant branch (TP-AGB). We argue that other studies missed this detection because of coarser age binnings.  \cite{2005MNRAS.362..799M} and \cite{2010ApJ...724.1030G} models predict the presence of a populated TP-AGB at  $\sim$0.6 Gyr, with a correspondingly very red integrated color, at variance with the data; \cite{2003MNRAS.344.1000B} and \cite{2009ApJ...699..486C} models run within the error bars at all ages. The discrepancy between the synthetic colors of  \cite{2005MNRAS.362..799M} models and the average colors of MC clusters results from the now obsolete age scale adopted.  Finally, our finding that the TP-AGB phase appears to develop between $\sim0.6-1$\,Gyr is dependent on the adopted age scale for the clusters and may have important implications for stellar evolution.
 \end{abstract}


\keywords{Galaxies: evolution, galaxies: stellar populations, stars: evolution}



\section{Introduction}\label{intro}

Evolutionary population synthesis (EPS) models have ubiquitous applications in extragalactic astronomy, allowing us to estimate ages, stellar masses, star formation rates and histories, and metallicities of galaxies as a function of redshift (e.g. \citealt{2011spug.book.....G}). Yet despite their importance, EPS models are far from perfect. Stellar evolution theory provides the backbone for EPS and as such, poorly understood aspects of the theory can have sizable effects on the models. For this reason, it is vital to test and calibrate both stellar and EPS models. 

A notoriously problematic stellar phase is the thermally pulsing asymptotic giant branch (TP-AGB). This is a prominent phase of double shell burning of helium and hydrogen in stars of low and intermediate-mass 
($\sim$ 1-10 M$_{\odot}$), when stars reach their highest luminosities, synthesize $p-$ capture and $n-$capture elements, and eventually shed in a wind almost all their hydrogen-rich envelope before evolving to their final white dwarf stage (\citealt{1983ARA&A..21..271I}). TP-AGB stars are red, cool giants and therefore their impact on synthetic spectra and colors is important at near-infrared (near-IR) wavelengths. In turns, such effect scales with the amount of radiative energy released during the TP-AGB phase, hence on the amount of nuclear fuel that is burned in the two shells.

Being red (super)giants, TP-AGB stars lose mass at high rates, until the hydrogen-rich envelope is lost and stars run out of fuel.
Thus, the energetic outcome of TP-AGB stars critically depends on mass loss, a process that is physically poorly understood while empirical estimates of mass loss rates are affected by large uncertainties. It is therefore no surprise if different EPS models 
offer discrepant estimates of the TP-AGB contribution (e.g., \citealt{1998MNRAS.300..872M};~\citealt{2003MNRAS.344.1000B};~\citealt{2005MNRAS.362..799M}; \citealt{2007A&A...469..239M}; 
\citealt{2009ApJ...699..486C}).

\cite{1986ASSL..122..195R} first introduced the idea of comparing ``simple stellar population'' (SSP) models (models for single age, single metallicity stellar population)
 with star cluster data in order to calibrate the SSPs with these `simplest' stellar populations. The uniformity of ages and metallicities of the stars and the lack of internal reddening in the star clusters allow a direct comparison with SSP models. This approach worked very well at optical wavelengths, leading to relatively well calibrated models (e.g. Maraston et al. 2003; Thomas et al. 2003; Maraston 2005). Most calibrations are made in integrated light rather than using resolved color-magnitude diagrams (CMDs) since the goal application of the models is to understand the {\it integrated} light from distant galaxies. 

\subsection{Historical background}\label{history}

The study of the TP-AGB contribution to the light of stellar population has a long history. The first estimate goes back over thirty years \citep{1981AnPh....6...87R} and was based on the fuel consumption theorem, thus predicting that the TP-AGB contribution to the bolometric light of an SSP could approach or even exceed $\sim$50\% at intermediate ages, from $\sim$ 40 Myr to a few Gyr. This result was based on the particular choice for the mass loss parameterization during the TP-AGB phase that had been adopted by \cite{1981A&A....94..175R}, which sounded reasonable at the time. Indeed, having placed a parameter $\eta$ in front of \cite{1975MSRSL...8..369R} mass loss formula, \cite{1976A&A....46..447F}  showed that a value $\eta\simeq 1/3$ for the mass loss during the red giant branch (RGB) phase was required to reproduce the broad properties of the horizontal branch of Galactic globular clusters. The same value of $\eta$ was also adequate to prevent evolving globular cluster stars from exceeding the RGB tip luminosity during their AGB phase, as demanded by the observations. Thus, \cite{1981A&A....94..175R}  limited their exploration of the parameter space to the range $1/3\le\eta\le 2/3$.  In support of the TP-AGB contribution predicted by  \cite{1981AnPh....6...87R} came the discovery that in many intermediate age globular clusters of the Magellanic Clouds (MCs) TP-AGB stars do indeed contribute a major fraction of the cluster light \citep{1983ApJ...266..105P}.  This description of the TP-AGB evolution had also other attractive aspects. For example, with full operation of the third dredge-up process through the whole initial mass (M$_{\rm i}$) range of TP-AGB stars (1 M$_{\odot}$$\lesssim$ M$_{\rm i}$ $\lesssim$ 8 M$_{\odot}$) the \textsuperscript{22}Ne neutron source appeared able of producing the s-process elements in the right amount and in near-solar proportions \citep{1978ApJ...220..980I}. Moreover, the core of the most massive TP-AGB stars could reach the Chandrasekhar limit, leading to a thermonuclear supernova explosion inside a still massive hydrogen-rich envelope 
[a SN of Type 1 and 1/2, as it was dubbed in \cite{1983ARA&A..21..271I}].

However, at nearly the same time it became apparent that TP-AGB evolutionary models such as those of \cite{1981A&A....94..175R} were dramatically over-predicting the number of TP-AGB stars brighter than 
M$_{\rm bol}\;\simeq -6$ compared to star counts in the MCs  \citep{1981ApJ...249..481C}, a discrepancy that was called the ``AGB star mystery''. Reluctant to abandon 
the attractive aspects of this theoretical scenario, \cite{1983ARA&A..21..271I} explored the possibility that in massive  TP-AGB stars the  stellar wind could be optically thick even in the $K$ band, thus making TP-AGB stars to appear dimmer than they are, but no clear theoretical solution to the ``AGB  star mystery'' emerged during the rest of the 'eighties. Therefore, a major luminosity contribution by TP-AGB stars for populations in the age range $\sim 30$ Myr to $\sim 1$ Gyr was still considered possible, but it was also clear that a proper calibration
of  the models with observations was indispensable. Ideally, the calibrators ought to be resolved stellar systems of which one could measure the age, count individual TP-AGB stars and measure their contribution to the various photometric bands.
Thus, star clusters in the MCs appeared to be the most suitable calibrators \citep{1986ASSL..122..195R}. They are located close enough for deep CMD studies using both HST and ground-based telescopes and, unlike those in the Milky Way, the MCs' clusters span ages in the range 0.3$\lesssim$ age [Gyr] $\lesssim$3, when the TP-AGB contribution is expected to peak. 
 
On the theoretical side, the breakthrough  indicating a way of solving the ``AGB mystery'' finally came from  models by \cite{1991A&A...244L..43B}. In relatively massive TP-AGB stars  (i.e., those with initial mass M$_{\rm i}$ $\gtrsim$ 3 M$_{\odot}$) the base of the convective envelope is hot enough to support Hydrogen burning (hot bottom burning); \cite{1991A&A...244L..43B} discovered that under
these conditions, the standard core mass - luminosity relation  [known as the Paczy\'nski's core mass-luminosity relation \citep{1970AcA....20...47P}], assumed in \cite{1981A&A....94..175R}, does not hold. Instead, the luminosity increases much more rapidly with core mass, possibly leading to a prompt envelope ejection, hence an early termination of the TP-AGB phase, and a drastic reduction in both the number of bright TP-AGB stars and their contribution to the integrated light of stellar populations. 
Since this applies only when hot bottom burning is active, the TP-AGB phase is prematurely aborted only in the high mass range,  which is however difficult to pinpoint, since  modeling the stellar evolution through this phase is particularly uncertain.

These studies made it clear that the TP-AGB evolution critically depends on several parameters that could not be predicted from first principles, and therefore a suitable observational calibration was needed. These parameters included the mixing length for the envelope convection (controlling the efficiency of the hot bottom burning process, hence the break-down of  Paczy\'nski's relation), the strength of stellar winds and their dependence on stellar basic parameters, and the efficiency of the third dredge up process (controlling the growth rate of the stellar core).  
\subsection{Model calibration with Magellanic Cloud star clusters.}\label{MCs}
Rather than trying to calibrate each of the above parameters separately, the pioneer work of Maraston (1998) aimed at calibrating the {\it total} TP-AGB {\it fuel consumption}, i.e., the AGB contribution to the integrated bolometric light, as well as the optical and near-IR colors of synthetic stellar populations. To this end, a sample of Small and Large Magellanic Cloud (SMC/LMC) star clusters were used  and predictions for population models with fixed (roughly solar) metallicity were obtained. 
  \cite{2005MNRAS.362..799M} extended this calibration to the calculation of full theoretical  spectral energy distributions of population models by including empirical Carbon and Oxygen star spectra from \cite{2000A&AS..146..217L} and \cite{2002A&A...393..167L}. The energy partition between carbon-rich (C type) and oxygen-rich (M) type TP-AGB stars as a function of metallicity was assigned by following the theoretical arguments of \cite{1981A&A....94..175R}. 


The semi-empirical approach of Maraston uses MCs' clusters as a basis to fix the energetic and colors of population models featuring TP-AGB stars. Hence, the resulting models will depend on the calibrating data and some relevant quantities associated to these, namely:
i) the {\it ages} assigned to the clusters, which fix the epoch at which the TP-AGB develops; and
ii) the integrated photometry of clusters, which determine the colors of the integrated models.

Related to the above, the set of tracks, on which the age determination is based, is key. As well known, convective overshooting affects the luminosity of the turnoff and other details of post-MS evolution. Because of the longer nuclear timescales of stars in tracks including overshooting, cluster ages derived with the latter tend to be older than those obtained with classical, i.e. non overshooting, 
 tracks (e.g., \citealt{1995A&A...298...87G},  \citealt{1995MNRAS.272..391F},  \citealt{2004ApJ...608..772F}, \citealt{2006ApJ...646..939M}). 
 The exact age shift depends on the assumed overshooting, but it is of the order of $\sim$0.3-0.5 Gyr \citep{2004ApJ...608..772F}. 
 Another important aspect are the cluster luminosities and colors that can be used on an individual basis, or averaged out between objects.

\cite{1998MNRAS.300..872M, 2005MNRAS.362..799M} derived the TP-AGB fuel consumption as a function of cluster age, adopting the age calibration by \cite{1990ApJ...352...96F} which is based on classical, no-overshooting stellar models. This choice was taken for consistency with the input stellar tracks of the models. For assessing the bolometric contribution of TP-AGB stars, the data for individual clusters were averaged in bins of representative cluster ages in order to minimize stochastic fluctuations among clusters with similar ages which are due to the short duration  of the TP-AGB phase. The clusters' photometry was taken from the databases available at the time, namely \cite{1981A&AS...46...79V} for the optical, and  \cite{1983ApJ...266..105P} for the near-IR. In \cite{2005MNRAS.362..799M} the $R,I$ photometry was added 
(from \citealt{2006MNRAS.369..697G}) in order to cover the whole SED for the calibration. Clusters were not stacked to determine average colors in age bins.

 \cite{2009ApJ...699..486C} and \cite{2010ApJ...712..833C} adopted the same approach to calibrate their models, but used more recent results for MCs clusters in terms of luminosities, colors and ages, from \cite{2006AJ....132..781P}, 
\cite{2008MNRAS.385.1535P} and also {\it averaged} cluster colors from \cite{2004ApJ...611..270G}.  By adopting systematically older cluster ages their calibration postpones the impact of the TP-AGB phase on integrated colors and it is found less prominent compared to Maraston's models.

 
The use of average instead of individual colors like in the \cite{2005MNRAS.362..799M} models already explains part of the discrepancy highlighted by \cite{2010ApJ...712..833C}, but the method used for averaging colors and binning in age is worth further investigation. 
Our goal here is: i) to investigate the origin of the age discrepancy; and ii) to analyze the robustness of average data by playing with the age binning.

To this end, we consider a sample of 43 LMC and SMC clusters in different age ranges of up to $\sim 3$\,Gyr old.  

The paper is organized as follows. In section~\ref{data}, we present the data and describe the method adopted to obtain average magnitudes and colors and age binning. In section~\ref{results}, we analyze the results. Finally, in section~\ref{discussion} we discuss our findings, and present our conclusions.
\section{Sample Selection}\label{data}
The TP-AGB phase mostly affects the near-IR light of stellar populations of intermediate age, and its effects are best mapped in optical-infrared color combinations. To calibrate for this effect  on EPS models, we need
 stellar populations templates of independently known age, for which optical-to-near-IR colors are available,  and of sufficient size to ensure they include a fair number of TP-AGB stars 
 in the age interval in which these stars contribute an important fraction of the light. Therefore, for this goal,  we have compiled  from the literature a sample of MCs' clusters following these selection criteria:

\par\noindent
$\bullet$ Ages between $\sim$ 50 Myr and $\sim$ 3 Gyr, where ages were derived from isochrone fitting to 
CMDs\footnote{For reasons given below we allow just one exception to this criterion, the cluster NGC~2107, for which the age is derived from integrated colors.}

\par\noindent
$\bullet$ Availability in the literature of integrated optical and infrared luminosities and colors measured over a nearly constant aperture. Mandatory data include $V$- and $K$-band integrated magnitudes.

The latter is very important since when constructing colors it is substantial to use magnitudes measured over identical apertures, to ensure that the same fraction of the total light is sampled in both bands. The size of the aperture is however critical, as one would like it to include  \textit{all and only} the cluster population. For example, for an AGB bright star located far from the cluster's center, but still a member of the cluster,
 a big aperture should be used. However, if the AGB star does not belong to the cluster,  a smaller aperture would yield a correct estimate of the integrated color. Unfortunately, for many interesting MCs' clusters optical and infrared magnitudes {\it measured over the same apertures} are not yet available. We will return to aperture effects in Section~\ref{discussion}.

\subsection {Clusters' Photometry}
Following the above criteria, we selected MCs' clusters  from three different sources: 
\cite{1983ApJ...266..105P}, \citet[2008]{2006AJ....132..781P}, and
\cite{2006MNRAS.369..697G}.

\cite{2006AJ....132..781P}  used the 2MASS survey\footnote{Two Micron All Sky Survey} (\citealt{2006AJ....131.1163S}) to derive near-IR integrated 
magnitudes and colors for a sample of 75 MCs' star clusters based on such homogeneous, photometrically calibrated dataset. 
Integrated magnitudes were determined from curves of growth as functions of the aperture after a centering on the cluster's light. Their methodology of integrated light measurement and some caveats regarding field contamination is discussed in \cite{2010A&A...510A..19L}, and we shall return on this point in Section~\ref{field_subtraction}.
Most of the clusters in this sample have age and metallicity estimates from CMDs and cover an age range between $\sim$10 Myr to 13 Gyr old. These ages were taken from various sources in the literature and their  estimates are based on tracks with some convective overshooting and a mix of techniques.

\cite{2008MNRAS.385.1535P} added the IR magnitudes of nine more objects, and combined the  2MASS data with $B$-band and $V$-band photometry from the literature,  to yield integrated optical-infrared colors  typically measured over a 60\arcsec aperture. 

\cite{1983ApJ...266..105P} data, which were used to perform the calibration of \cite{2005MNRAS.362..799M}, include $J$-, $H$- and $K$-band photometry of 84 MCs' clusters spanning a very broad range of ages. The authors provide a $(V-K)$ color  for most but not all of these clusters, due to difficulties in matching the aperture between IR and optical photometry. 

\cite{2006MNRAS.369..697G} provide integrated-light photometry in the $VRI$ bands for a sample of 28 bright star clusters selected from \cite{1990ApJ...352...96F} and use near-IR photometry from the same source. The sample includes clusters with ages between 50 Myr and 7 Gyr, focusing in particular on clusters in the age range of 0.3 $\leq$ age $\leq$ 2 Gyr where the TP-AGB phase should be relevant. One of the main aims of this work was to include the $V-R$~and $V-I$ colors for MCs' clusters,  such as to allow the calibration of EPS  models over abetter sampled spectral energy distribution.   
 \cite{2006MNRAS.369..697G} give a range of apertures, for our case we used those for an aperture of 60$\arcsec$ in consistence with the rest of the data. 

In summary, for most clusters in our sample, integrated magnitudes and colors were measured over apertures of the order of $\sim 60''$ (see Table \ref{data_clusters}). 
We did not use the data from \cite{2006ApJ...646..939M} since their near-IR photometry for 19 clusters was obtained with an aperture of $\sim$90$\arcsec$ which remains unmatched to the optical wavelengths. 

\subsection{Age determinations}\label{indicators}
Accurate, homogeneous age determinations of individual clusters are of key importance for calibration purposes. For MCs' clusters, over the last decades ages have been  obtained adopting three methods:
 1) isochrone fitting to the CMDs of resolved stars; 2) the SWB type  classification \citep{1980ApJ...239..803S}; and 3) the so-called $s$-parameter classification \citep{1985ApJ...299..211E}. The latter two methods rely on  clusters'  integrated colors which vary systematically with the cluster age and metallicity. The relation between these indicators and the actual cluster age was derived by means of a calibration on a few objects for which the age could be determined from the direct analysis of the CMD.

The SWB  type  provides a classification scheme for rich  
clusters into seven types, on the basis of two reddening-free parameters derived from integrated
{\it uvgr} photometry. This sequence is interpreted in terms of increasing age and decreasing metallicity. 
\cite{1998MNRAS.300..872M, 2005MNRAS.362..799M} adopted the SWB type - age relation in \cite{1990ApJ...352...96F} [see their Table 3 and Figure 2],  
 who surveyed for AGB stars in 39 MCs' clusters.
 Thus, the ages associated to each SWB type are: Type I: 0.013 Gyr, Type II: 0.04 Gyr, Type III: 0.12 Gyr, Type IV: 0.37 Gyr, Type V: 1.1 Gyr, Type VI: 3.3 Gyr and Type VII: 10 Gyr. 
 
Note that this relation was calibrated on the relatively few CMD-based cluster ages and stellar models that were available at the time. A calibration performed today would yield different results: for example, two of the three clusters with SWB Type IV  in \cite{1990ApJ...352...96F} data base are now confirmed  with an age of $\sim$1 Gyr (see e.g. \citealt{2006AJ....132..781P}), and actually did not fit well into the SWB Type-age relation adopted by \cite{1990ApJ...352...96F}. We notice that \cite{1990ApJ...352...96F}  assumed a distance modulus of 18.3 for the LMC, which is $\sim$ 0.2 mag closer than currently adopted value. 
 In addition, \cite{1990ApJ...352...96F} adopted a relation between SWB type and age that underestimates the age of the calibrating SWB types III and IV clusters. 
 An example of this is that three SWB type IV clusters that  in  \cite{1990ApJ...352...96F} have ages of $\sim 0.37$ Gyr, namely NGC 152, NGC 1987 and NGC 2107, actually have ages of 1.4 Gyr, 1.08 Gyr and 0.62 Gyr, respectively, see Table \ref{data_clusters}, 
 as derived from CMDs by \cite{2001AJ....122..220C}, \cite{2009A&A...497..755M} and \cite {2008MNRAS.385.1535P}. 

The $s$-parameter  is a photometric age indicator based on integrated $UBV$ photometry, constructed as a curvilinear coordinate along the two-color locus occupied by the clusters. Being a continuous function of age, it represents a refinement of the SWB classification. The calibration of the $s$-parameter versus age has been modified several times depending on the set of template clusters and isochrones used to fit the CMD of them.
Particularly relevant to this - as already mentioned - is whether core convective overshooting is included or not in the stellar evolutionary models. Cluster ages are older at fixed value of $s$ in the former case.  Relations used in the literature include \cite{1988AJ.....96.1383E} for canonical models; \cite{1996A&AS..117..113G} for models with overshooting;  and \cite{2008MNRAS.385.1535P} for the same models with overshooting, but including newer age determination for a few calibrating clusters.


All these methods which are based - directly or indirectly - on the CMD age-dating method rely on stellar evolution models. There is good agreement between different stellar models, but - as already mentioned - some discrepancies remain regarding the treatment of convective core overshooting in stars more massive than $\sim$1 M$_{\odot}$ (see \citealt{2005ARA&A..43..387G} for a review). 
Overshooting increases the amount of fuel available for core hydrogen burning and hence prolongs the MS lifetime, thus increasing the age derived for  the clusters compared to the use of no-overshooting models. 
The extent to which overshooting operates remains conjectural and is parameterized in different ways, usually after calibration with observations. In Galactic open clusters  a certain amount of overshooting is required (see e.g. \citealt{1992ApJ...387..320C}).
While the tests performed on young MC star clusters ($\lesssim$1 Gyr old) yield to contradictory (e.g. \citealt{1999AJ....118.2839T}; \citealt{2006ApJS..162..375V}; \citealt{2011spug.book.....G})
or inconclusive (e.g. \citealt{2003AJ....125.3111B}) results,  those on intermediate-age cluster (in the 1-3 Gyr range) present good agreement using moderate overshooting prescriptions
 (e.g. \citealt{2003AJ....125..754W}). In summary, ages derived using overshooting models are typically $\sim 30\%$ older than those using no-overshooting models.
   
Finally, one additional complication concerning the ages of MCs' clusters comes from finding evidence of  prolonged star formation in several of them as revealed by MS turnoffs that are exceedingly broad compared to photometric errors (e.g., \citealt{2003AJ....125..770B}; \citealt{2007MNRAS.379..151M}; \citealt{2008ApJ...681L..17M}; \citealt{2008AJ....136.1703G}; \citealt{2009A&A...497..755M}; \citealt{2011ApJ...737....3G}). Hence, there seems to be solid indication that many MCs' clusters host either multiple stellar populations or had prolonged star formation, on timescales of some $10^8$   years.
Still, this does not affect our age assignment since we stack clusters in broad age bins, though from a conceptual point of view we use these clusters as SSP templates while strictly speaking they are not.

The age for virtually all clusters in our sample comes from the interpretation of the CMD, albeit with different sets of tracks, and/or in different colors, and/or with data down to different depths and
therefore the uncertainty on individual ages varies through the data set.  However, in order to gain statistical significance we group the clusters in age bins and  the accuracy of individual ages is not important as long as a cluster remains in the assigned bin. 

Finally, we stress that we do not particularly favor CMD-based ages derived from isochrones with or without overshooting. We simply use the most updated CMD-based ages in order to age-date in the most homogeneous way the clusters. In the Maraston (2013) models (see Section \ref{integrated}) the onset age of the TP-AGB has been modified to match the present age-calibration. More comments will follow in that paper. 

\subsection{Field subtraction effects on globular cluster colors}\label{field_subtraction}
As discussed in \cite{2010A&A...510A..19L}, the $K$-band magnitudes of \cite{2006AJ....132..781P} are fainter by approximately one magnitude compared to those reported  by \cite{2006ApJ...646..939M} for NGC 1806 and NGC 2162  having used  a similar aperture whereas the $(J-K)$ colors are bluer. These effects could be either due to an overestimated LMC field decontamination if  \cite{2006AJ....132..781P}  removed the reddest stars in the aperture, or to an underestimate  of the field contamination in the case of \cite{2006ApJ...646..939M}. \cite{2010A&A...510A..19L} checked these possibilities in the case of NGC 2162, where the $K$-band light of the globular cluster is dominated by a single carbon star. By adding its $K$-band luminosity, taken from the 2MASS catalogue, to the integrated luminosity given by  \cite{2006AJ....132..781P}, they obtain a $K$-band luminosity in much better agreement with \cite{2006ApJ...646..939M}. Field carbon stars are not very common  at the LMC location of this cluster, thus there is a high probability that this star is a cluster member. Therefore, \cite{2010A&A...510A..19L} concluded that the large color gradients (i.e. differences of up to 0.2 mag in $(J-K)$ colors depending on the aperture) present in the work of  \cite{2006AJ....132..781P}  and their fainter $K$-band magnitudes may be mostly due to their over-subtraction of the LMC field star contribution for these two clusters. 
A similar conclusion is reached by \cite{2013arXiv1304.3499S} who found some discrepancies with the photometry from some objects younger
 than $\sim$100 Myr that they have in common with \cite{2006AJ....132..781P}. 
 Without doubt,  field subtraction is a very delicate operation, particularly in the near-IR where short living bright stars are important.

\subsection{Final sample}\label{sample}
To construct our final sample we merged the \cite{2008MNRAS.385.1535P} and the \cite{1983ApJ...266..105P} sample, adopting \cite{2008MNRAS.385.1535P} photometry and most recent ages from CMDs 
(see references in Table \ref{data_clusters}). 
We checked the literature for updated age determinations, based on the analysis of the CMDs, and singled out from the final sample those clusters with age between 0.05 and 3 Gyr.  
 The ages adopted in Table~\ref{data_clusters} imply a systematic shift with respect to the \cite{2005MNRAS.362..799M} calibration,  which was based on the different set of age indicators that were available at that time (see Section \ref{indicators}).

Following the selection criteria mentioned at the beginning of this Section, our final set consists of 43 MC star clusters that are listed in Table \ref{data_clusters} which gives the de-reddened $V_{0}$ magnitudes; the de-reddened colors: $(V-R)_{0}$, $(V-I)_{0}$, $(V-J)_{0}$, $(V-H)_{0}$ and $(V-K)_{0}$; the $V$-band extinction;  the aperture diameters, ages, and corresponding references. Reddening has been taken from 
the Magellanic Clouds Photometric Survey.\footnote{http://djuma.as.arizona.edu/$\sim$dennis/smcext.html}
The $V$-band magnitude was taken from  \cite{2006MNRAS.369..697G} for the 25 clusters we have in common and for 
the rest we took it from \cite{2008MNRAS.385.1535P} or from other authors when not given by the latter (see references in Table \ref{data_clusters}).
  
As mentioned above, we carefully took magnitudes in all bands obtained with similar apertures. 
Since in many cases the $V$-band photometry is originally from  \cite{1981A&AS...46...79V}, largely obtained with $\sim$60$\arcsec$ apertures, most of our data refer to this value.
Table \ref{data_clusters} also includes optical colors for those clusters which appear also in the  \cite{2006MNRAS.369..697G} sample.

As already mentioned, the ages listed in Table \ref{data_clusters}  have been derived by different authors, adopting different sets of stellar evolutionary sequences and using
photometric data of various qualities. Hence, such ages cannot be qualified as homogeneous, and this remains still a concern for the calibration of EPS models. Notice that the age determination is also dependent on adopted distance modulus and reddening,
on the metallicity and in general on the method used to interpret the stellar distribution on the CMD.
Different options for this ingredients introduce a systematic effect on the cluster ages which is very difficult to  correct for. Although the shortage of AGB stars in individual clusters forces us to stack them in age bins, nonetheless, gathering strictly homogeneous photometric data for most MCs' clusters and homogeneously derived ages for them remains a major unfulfilled need for many astrophysical applications.

\subsection{Age binning}\label{binning}
One of the most important sources of error when calibrating EPS models using clusters is the
rareness of AGB stars with individual clusters containing at most just a few such stars. With the aim to overcome stochastic effects and increase our statistics we then stacked the MCs' clusters in our sample in different age bins. This is non trivial since the selection of these age intervals can impact on the resulting calibration.  The AGB phase transition may give a precise signature at a specific age but a too broad or misplaced  age binning could dilute this feature, averaging the characteristics over a wide age range.  

After assigning ages to each cluster in our sample as described in Section \ref{indicators}, 
we performed a careful analysis testing different age binning. 
As a compromise between the need to finely describe the variation with age  of the cluster properties, and that of maximizing the statistical significance of the 
stacked clusters, we selected the following five bins where ages are in Gyr:
 age $<$ 0.3 (age1),  0.3 $\leq$ age  $\leq$ 0.9 (age2),   0.9 $<$ age $\leq$ 1.4 (age3),  1.4 $<$ age  $<$ 2 (age4), 
and 2 $\leq$ age  (age5) as reported in Table \ref{n_agb}.
This selection helps enhancing the $V$-band luminosity sampled at ages just 
below 1 Gyr and in our oldest bin. These are critical ages for the effect we aim at mapping, and the $V$-band luminosity is the best tracer of the sampled mass, while the IR is also very sensitive to the presence of just a few TP-AGB stars.

Table~\ref{n_agb} provides details about the stellar population sampling of each bin, such as total luminosities, mass, and expected number of TP-AGB stars for each million year duration of such phase. In the next section we discuss the statistical properties of the resulting {\it stacked clusters} also called {\it superclusters}. Throughout the whole paper we will show the individual and the stacked clusters used for our analysis. 

Finally, we stress the fact that we decided to include NGC 2107 even though is the only cluster without CMD age determination, in order to have more statistic in our second age bin (age2).  

%
\section{Results}\label{results}

\begin{figure*}
\begin{center}
\includegraphics[width=0.9\textwidth]{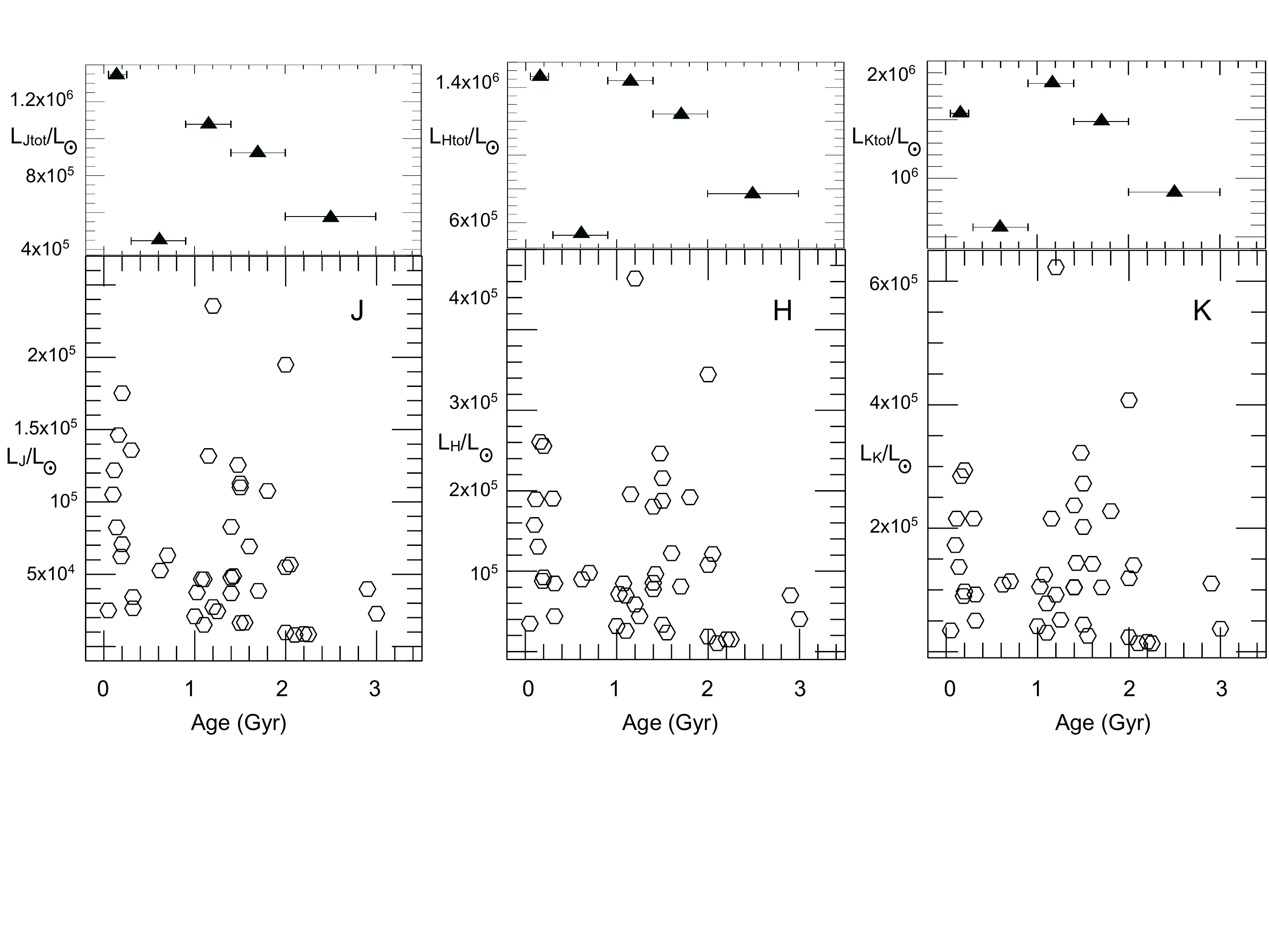}%
 \caption{Near-IR luminosities of all the MCs' star clusters in our sample are  shown with empty hexagons (lower panels). 
The cumulative luminosities of all the clusters in each age bin are shown as black triangles (upper panels)  where horizontal bars denote the age interval.} 
\label{lum_JHK}
\end{center}
\end{figure*}

\begin{figure*}
\begin{center}
\includegraphics[width=0.7\textwidth]{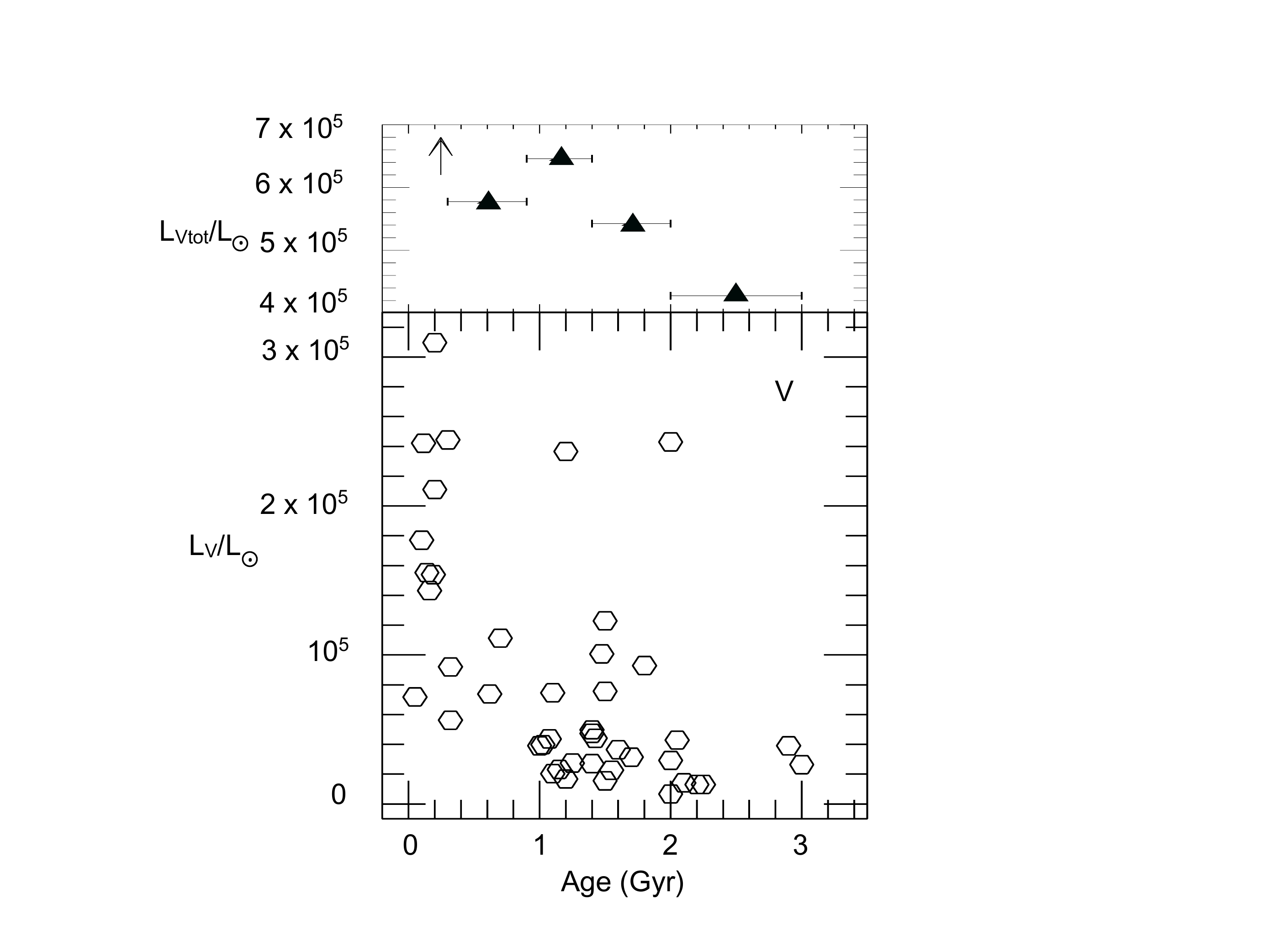}%
 \caption{The same as in Figure \ref{lum_JHK} but for the $V$-band luminosities of all the MCs' star clusters in our sample. For clarity purposes the luminosity in the first age bin (age1) is denoted by an arrow.}
\label{lum_V}
\end{center}
\end{figure*}

The integrated light in the near-IR and optical bands is crucial in order to address to which extent the data offer a fair sampling of TP-AGB stars. Hence, for each age bin we constructed the sampled luminosities and stellar mass as reported in Table~\ref{data_clusters}. This is shown in Figures \ref{lum_JHK} and \ref{lum_V}  where individual cluster luminosities in $JHK$ (Figure \ref{lum_JHK}) and in $V$ (Figure \ref{lum_V}) are denoted with empty hexagons and the total 
luminosities per age bin  are shown as solid black triangles.

\begin{figure*}
  \begin{center}
  \includegraphics[width=0.5\textwidth]{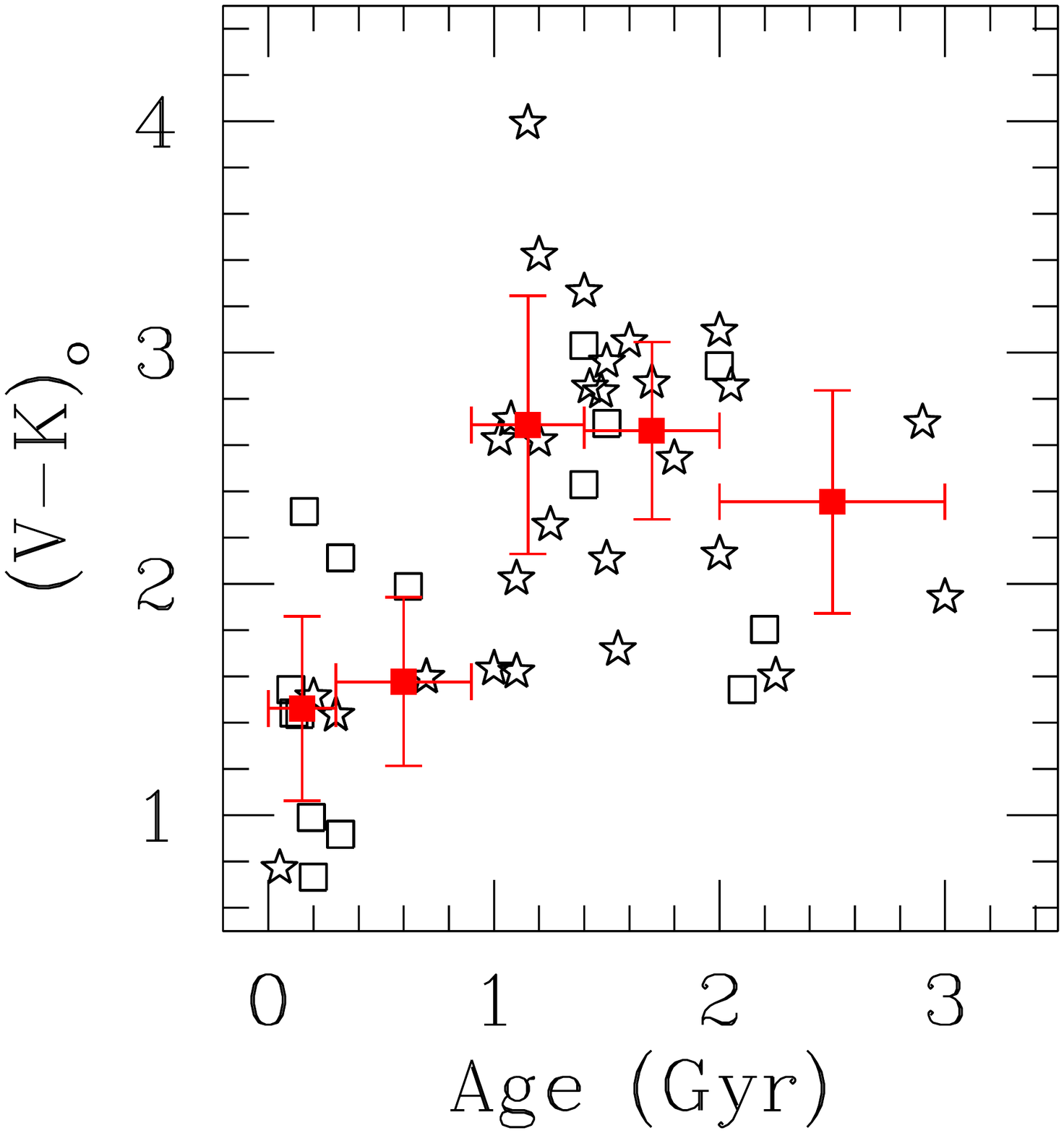}%
  \includegraphics[width=0.5\textwidth]{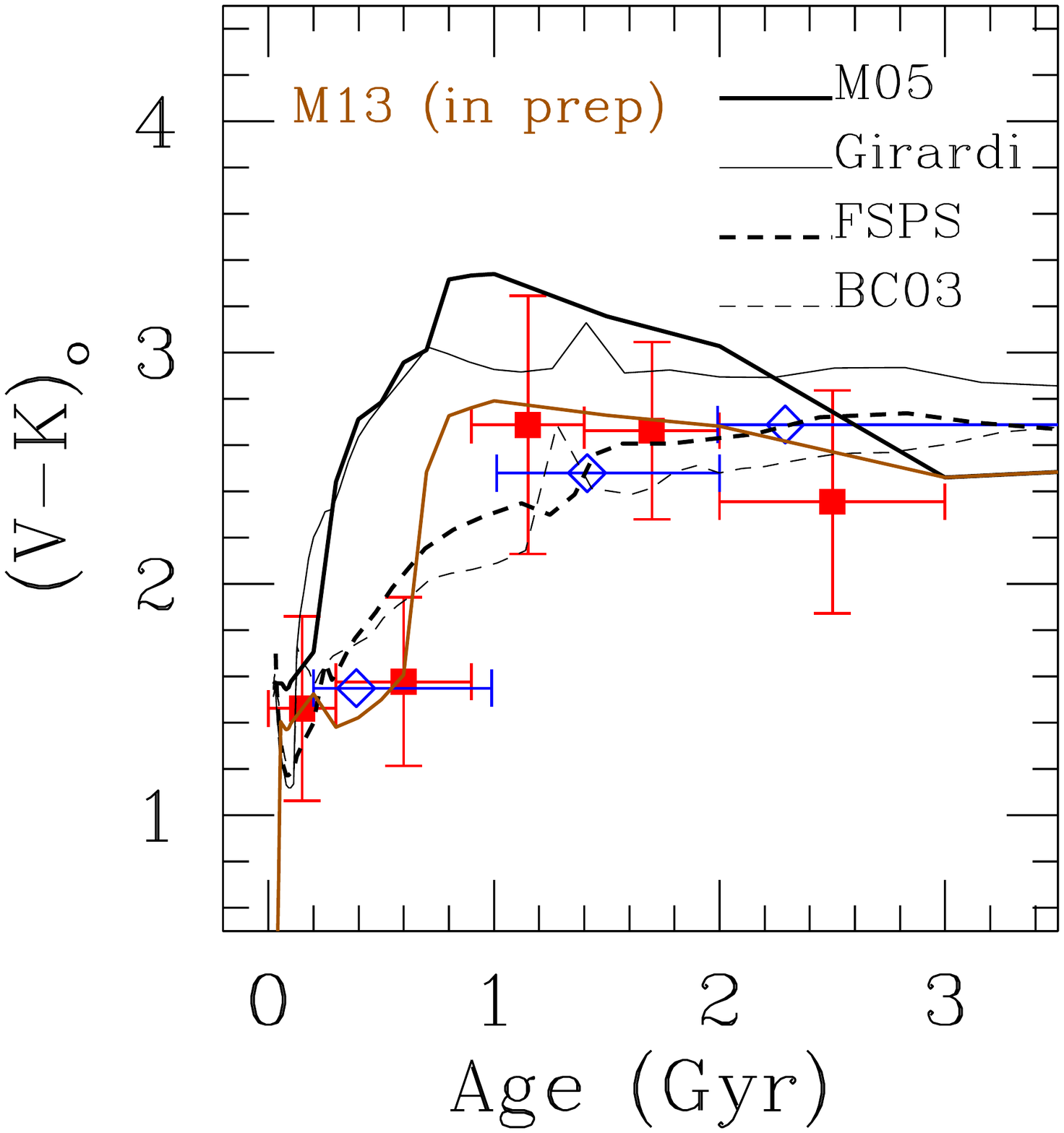} %
  \caption{The (V-K)$_{0}$ color as a function of age for the clusters in our sample. Left panel: empty stars and empty squares represent the individual clusters whereas  large red squares refer to the color for the stacked clusters in each age bin.
Vertical error bars represent the $V$-band luminosity weighted errors of the mean color
 and horizontal bars denote the size of each age bin. Right panel: Averaged data (from the left panel) with the models from M05, BC03, FSPS, Girardi et al. and M13 (in preparation) overlaid. 
  \cite{2008MNRAS.385.1535P}  superclusters are also reported as empty blue diamonds with the horizontal bars showing the age range they used.} 
 \label{color_models} 
 \end{center}
\end{figure*}
\begin{figure*}
 \begin{center}
  \includegraphics[width=0.5\textwidth]{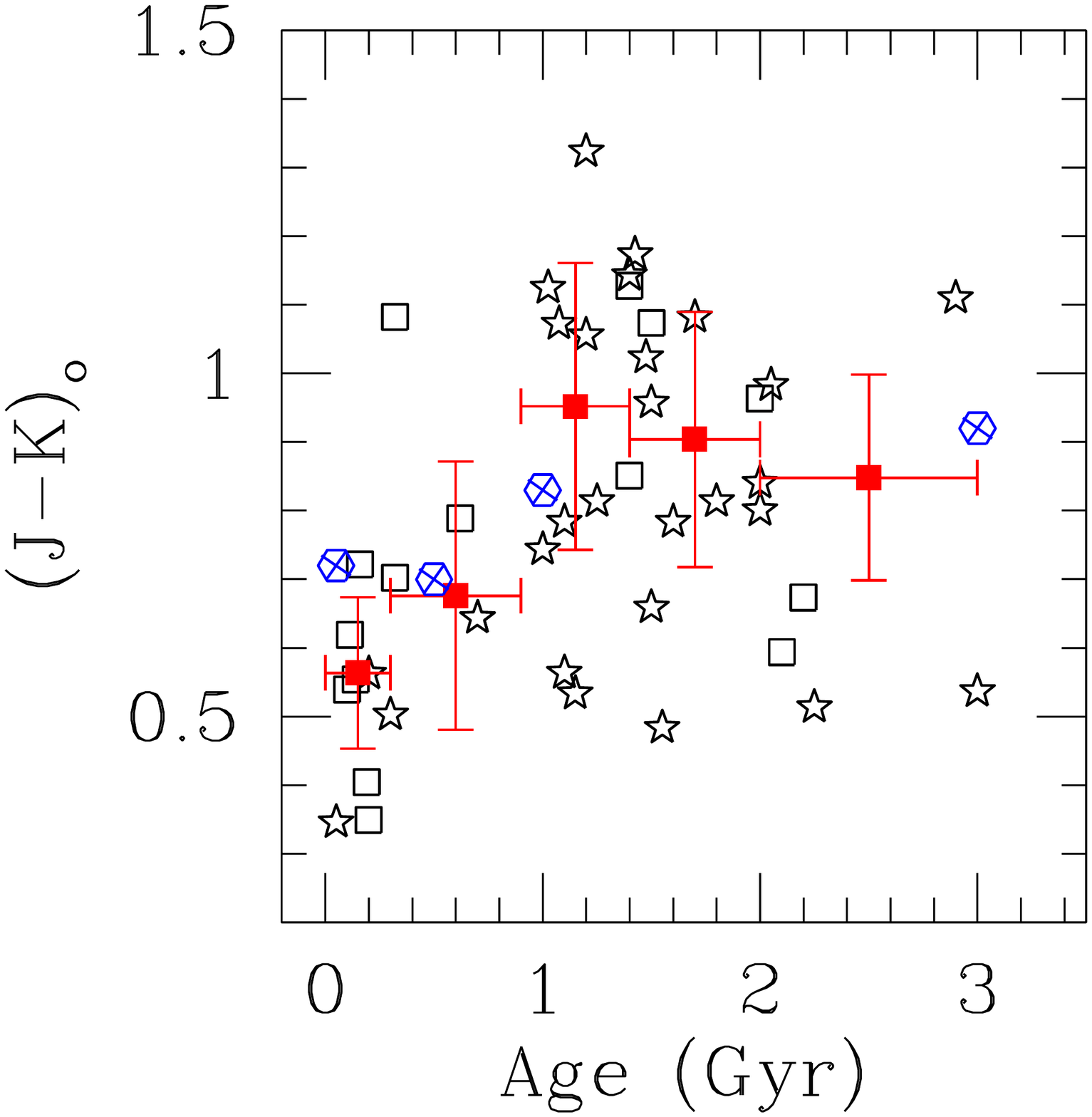}%
  \includegraphics[width=0.5\textwidth]{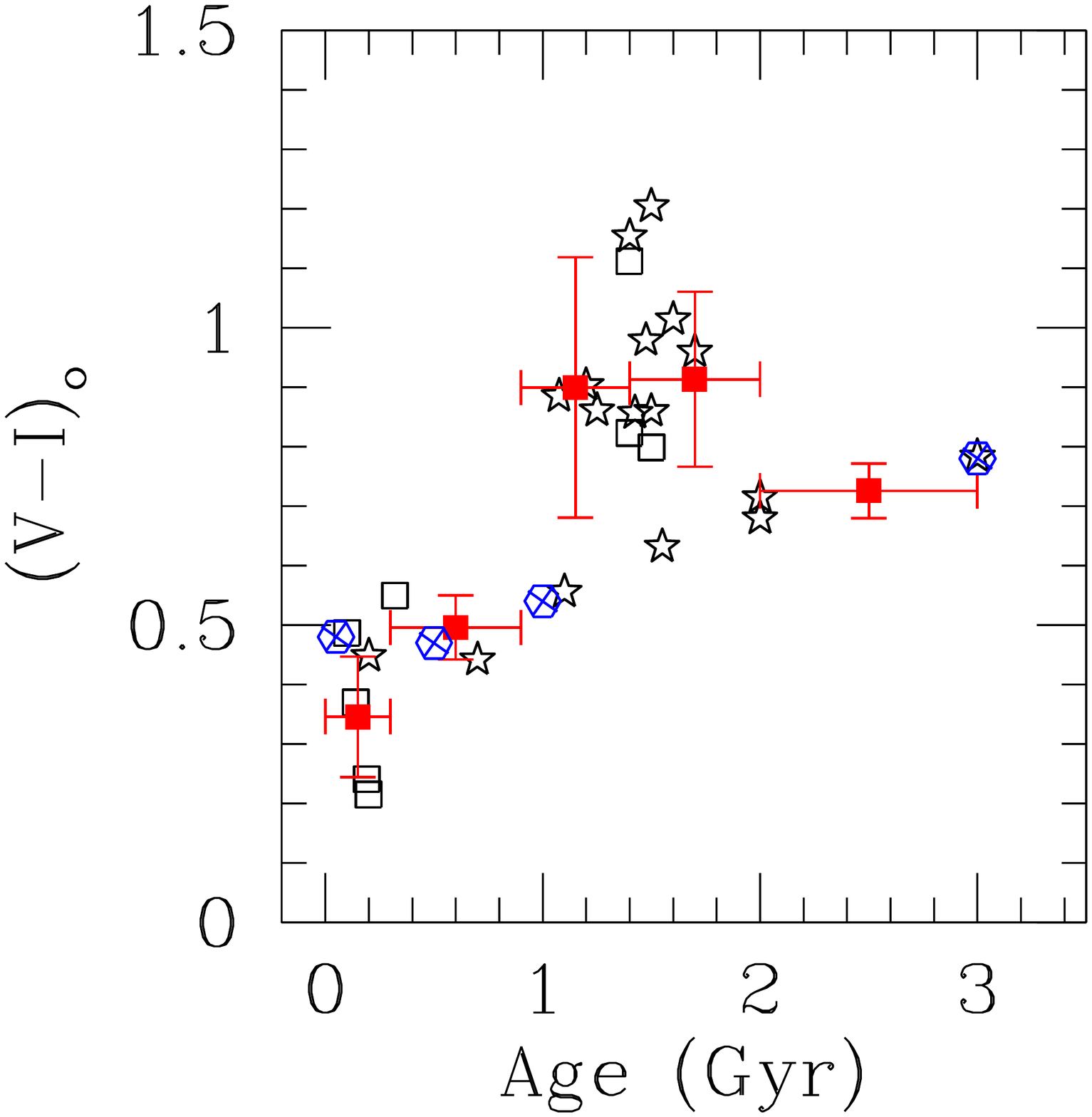} %
  \caption{The $(J-K)_{0}$ and $(V-I)_{0}$ colors as a function of age for the clusters in our sample. Empty stars and empty squares represent the individual clusters. Large red squares are colors for the stacked clusters in each age bin. Error bars have the same meaning as in Figure  \ref{color_models}. Superclusters from \cite{2004ApJ...611..270G} are also reported  as crossed-hexagons.} 
  \label{colors} 
 \end{center}
\end{figure*}
\begin{figure*}
  \begin{center}
  \includegraphics[width=0.85\textwidth]{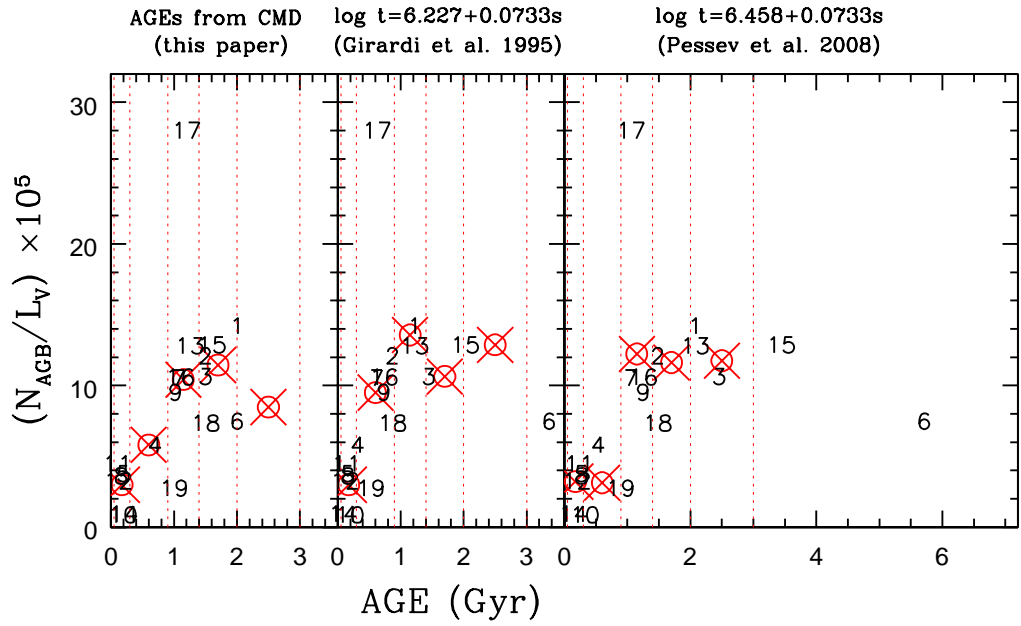}%
 \caption{The number of AGB stars brighter than the Tip of the RGB per unit solar $V$-band luminosity for clusters in \cite{2006ApJ...646..939M} for three options of the cluster age
determination, as labelled. The values for individual clusters are plotted as black numbers: for each cluster the same number is used
in the three panels.
Red circles show the same quantity for stacked clusters in each of our age bins The cluster number 17 is NGC 2209, with 4 AGB stars, in spite of its low luminosity; cluster number 6 is NGC 1978, for which the $s$ parameter indicates an age far in excess of that derived from the CMD by \cite{2009A&A...497..755M}.}
\label{nagb} 
 \end{center}
 \end{figure*}

\begin{figure*}
  \begin{center}
  \includegraphics[width=0.85\textwidth]{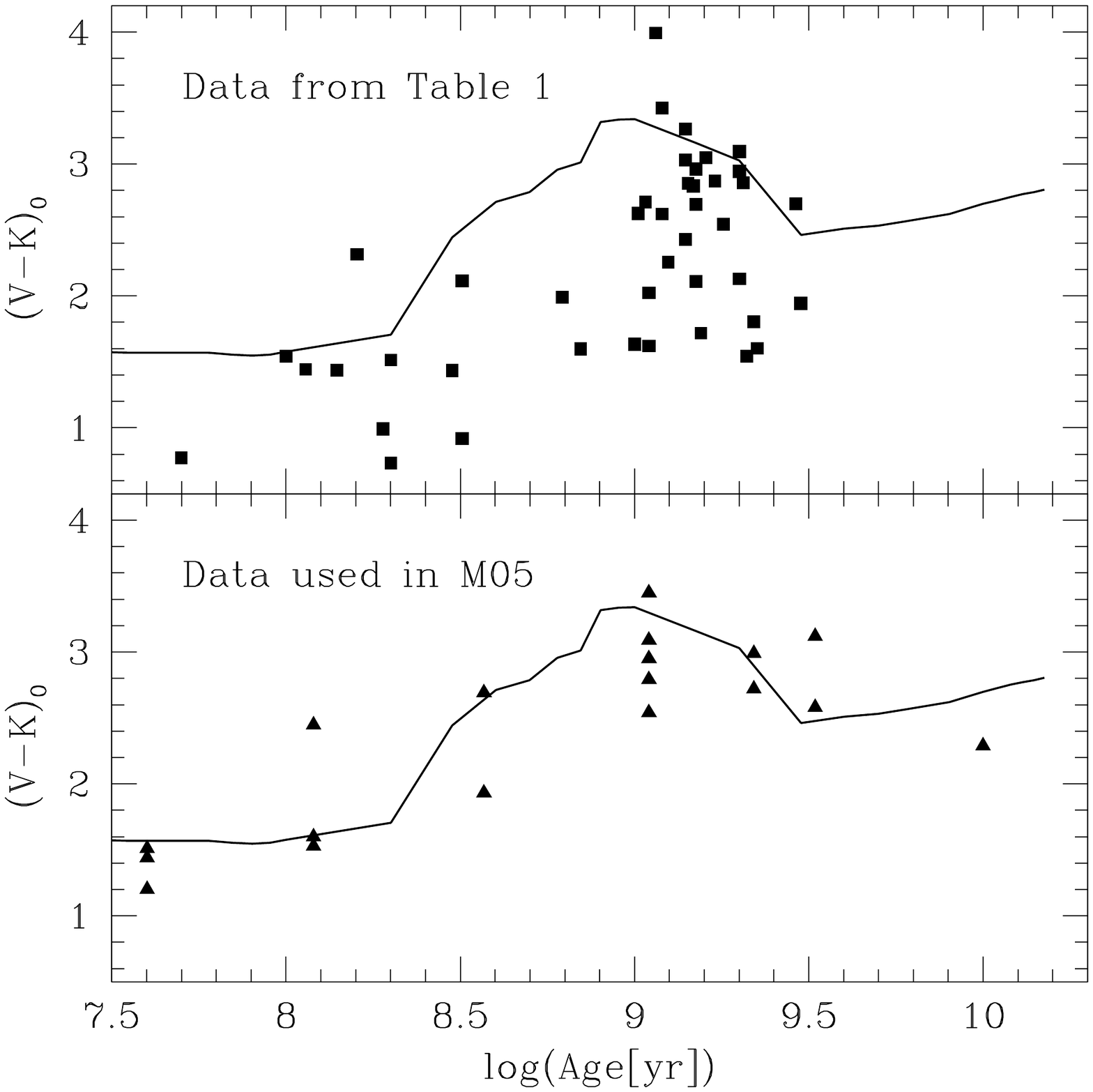} 
  \caption{Top panel: the  clusters' $(V-K)_0$ color as a function of age for all our clusters using the ages and colors from Table \ref{data_clusters}. Maraston (2005) models are overlapped. 
  Bottom panel: The clusters's $(V-K)_0$ color as a function of age from \cite{1990ApJ...352...96F} where ages  where derived from  SWB-type vs. age calibration and adopted in Maraston (2005), whose models are also overlapped.} 
 \label{two_panel} 
 \end{center}
\end{figure*}

Comparing Figure \ref{lum_JHK} to Figure \ref{lum_V} we notice that the luminosity sampled presents a deep minimum in the second age bin in the IR bands, while this is not the case in the V band.
This different behavior reflects the different sensitivity of the $V$ and the IR bands to the presence of a bright AGB,  the former being less sensitive to the cool AGB population. Thus, the low IR luminosity of bin age2 is not due to a low stellar mass sampled, otherwise also the luminosity in the $V$-band would present a deep minimum. Rather, an extended AGB is not yet fully developed in this age bin, while it is 
in the third  age bin. The luminosity sampled in bins age4 and age5 is low both in the IR and in the $V$ band, which is due to the older age of these clusters. 
However, their mass is actually higher than that in the other bins, as seen in Table \ref{n_agb}, where the cluster masses were calculated via the $M/L_{\rm V}$ ratio from the models of \cite{2005MNRAS.362..799M}. 

In order to evaluate the statistical significance of the stacked clusters, we estimated the number of stars in the TP-AGB phase which should be present given the sampled luminosity in the $V$-band (see \citealt{2011spug.book.....G}): 
\begin{equation}   
N_{\rm TP-AGB} = B(t) (L_{\rm bol}/L_{\rm V}) \times L_{\rm V} \times t_{\rm AGB} 
\end{equation}
where the specific evolutionary flux $B(t)$  increases from 0.8 to $1.9 \times 10^{-11}$ 
stars/$L_{\odot}$/year as a stellar population ages from 0.05 to 3 Gyr, the bolometric to visual luminosity ratio is $\sim 1.5$ in this age interval (from \citealt{2005MNRAS.362..799M}) and  t$_{\rm AGB}$ is the expected duration of the TP-AGB phase in years.  The clusters in our sample have been selected to encompass the age range in which the TP-AGB develops; therefore, 
t$_{\rm AGB}$ may vary considerably from the youngest to the oldest of our bins. 
As shown in \cite{2011spug.book.....G}, and also below, the lifetime of the TP-AGB phase is of the order of at most a few Myr.  In Table \ref{n_agb} we report  the number of TP-AGB stars expected in the stacked clusters if  t$_{\rm AGB} = 1$ Myr. We did not considered the clusters in the first age bin age1 for this estimate  since the TP-AGB transition did not occur yet at these young ages. 
The expected statistics seems acceptable, especially in bins age3 and age4.

%
\subsection{Integrated colors as function of age}\label{integrated}
In order to secure a proper calibration of EPS models, the trend of integrated colors with age is crucial.
We first consider the $(V-K)$ color for the clusters in our sample, as shown in Figure \ref{color_models}. 
The left panel of Figure \ref{color_models} displays each of the 43 individual  clusters with empty symbols while the stacked clusters in the five age bins are
shown as red filled squares. Error bars represent the $V$-band luminosity weighted standard deviation in each bin and the horizontal bars denote the age bins.
Star-like symbols represent the clusters with better age determination, based on more recent and deeper CMDs, as compared to other clusters, shown as empty squares, which are also based on CMD age determinations, but from shallower photometry.
The figure shows that clusters  become abruptly red in $(V-K)$ when aging from $\sim 0.6$ to $\sim 1$ {\rm Gyr}.
The $(V-K)$ color seems then to slightly decline at older ages, remaining close to $\simeq$ 2.5.
The abrupt reddening of the $(V-K)$ color is  attributed to the appearance of a well developed TP-AGB phase, which was earlier named ``the AGB phase transition" (\citealt{1981AnPh....6...87R}; \citealt{1986ASSL..122..195R}), whereas the subsequent decline of the TP-AGB contribution is partly compensated by the development of the RGB (the ``RGB phase transition'').

 The right panel of Figure \ref{color_models} shows our stacked clusters together with the same SSP models considered in the \cite{2010ApJ...712..833C} comparison, namely from \cite{2005MNRAS.362..799M} [M05; thick solid line], \cite{2003MNRAS.344.1000B} [BC03; light dashed line], \cite{2009ApJ...699..486C} [FSPS\footnote{FSPS models from \\ http://www.ucolick.org/$\sim$cconroy/FSPS.html.}; thick dashed line]. Additionally, we included the \cite{2010ApJ...724.1030G}\footnote{CMD tool version 2.4 from stev.oapd.inaf.it/cgi-bin/cmd.}  newest models (thin solid line).  Also over-plotted are the models from Maraston (2013, in preparation) [M13; solid golden line] that will be briefly discussed in Section~\ref{models}.  The blue empty diamonds represent the superclusters from \cite{2008MNRAS.385.1535P},
  with the horizontal bars showing the age range they used (see their Table 5).  
  
For all models shown in Figure \ref{color_models}  we adopted a metallicity of  Z$=$ 0.5 Z$_{\odot}$. 
Although there may be a weak dependence on age this would not affect the results. For example,  \cite{2010ApJ...712..833C}
adopted Z$=$ 0.52 Z$_{\odot}$  for ages $\le 1$ Gyr andZ$=$ 0.42 Z$_{\odot}$
for ages of $\sim 3$ Gyr, interpolating in between such ages, thus the metallicity differences
between our Figure \ref{color_models} and Figure 2 of \cite{2010ApJ...712..833C}  are negligible.

 At ages older than $\sim 1$ Gyr the various models are largely in agreement, with M05 and \cite{2010ApJ...724.1030G} models at the reddest end of  age bins age3 and age4, and BC03 and FSPS  at the bluest end of the error bars for the age bin age3. At older ages (bin age5) all models but \cite{2010ApJ...724.1030G} [that are still too red] converge to the observational value. 
A sizable difference is instead noticeable in the second age bin, age2,  at $t\sim 0.5$ Gyr, where M05 and \cite{2010ApJ...724.1030G} models are much redder than our averaged data
 whereas FSPS and BC03 models do not exhibit the sudden reddening in $(V-K)$ at about this age, that appears to be demanded by the data.
Notice that the colors of the stacked clusters in \cite{2008MNRAS.385.1535P} are very similar to ours, 
but having adopted a different age binning they did not find the abrupt transition of the $(V-K)$ color which we detect between our bins age2 and age3. This underlines the effect of binning when trying to detect the development of the TP-AGB among the clusters, and then use their color to calibrate the models.

In Figure \ref{colors} we show $(J-K)$ and $(V-I)$ colors as functions of age. Again, open symbols represent the individual clusters while the red filled squares are our stacked clusters. 
The blue empty crossed-hexagons here represent the stacked clusters from \cite{2004ApJ...611..270G}. 
As before, error bars represent the $V$-luminosity weighted standard deviation and horizontal bars are the age bins.
The same trend shown in Figure \ref{color_models} is evident in $(J-K)$ and $(V-I)$ too, in particular the
abrupt increase of the integrated color between bin age2 and bin age3. We interpret this effect as a confirmation that the TP-AGB develops between $\sim$ 0.5-0.6 and $\sim$ 1 Gyr.  
Note that less clusters are depicted in the right panel of Figure \ref{colors} showing  $(V-I)$ vs. age, since only 25 of the 43 clusters in our sample have $I$-band measurements.
As in the case of \cite{2008MNRAS.385.1535P},  the superclusters from  \cite{2004ApJ...611..270G} ``fade'' the AGB phase transition.

The (V-K) color provides the widest color baseline, hence it offers the best leverage to pinpoint the contribution of AGB stars which mostly contribute in the $K$-band whereas the $V$-band luminosity is 
dominated by main sequence stars. A comparison with models is therefore restricted here to this (V-K) color, with a thorough multicolor comparison being part of the M13 paper presently in preparation. 

We finally emphasize that our age bins do sample enough mass to be statistically significant for a duration of the TP-AGB phase around 1 Myr.  In other words, if this was the duration, in our age bin age2 we should sample
 $\sim$15 bright AGB stars, which seems adequate to determine the impact of this phase on the integrated color of the cluster, beyond statistical effects.
  Similarly massive is the supercluster in age bin age3, so that its red $(V-K)$ color ensures that TP-AGB stars give an important contribution to the light. 

\subsection{AGB star counts}
The direct way to map the development of the TP-AGB is clearly examining the individual stars on the CMDs of clusters with the appropriate ages.   
\cite{2006ApJ...646..939M} presented the $J,K$ CMDs of 19 intermediate age MCs' clusters, all of which are included in our sample and have ages determined from the analysis of the optical CMDs.
For each cluster, these authors give star counts on different evolutionary branches, in particular on the portion of the AGB brighter than the tip of the RGB. Being so bright, these AGB stars are very likely to be in the thermally pulsing phase.

Using these data we construct  Figure \ref{nagb}, which  shows the ratio between the number of AGB stars and the V-band luminosity, {N$_ {AGB}$}/{L$_{V}$}, for individual clusters (marked with numbers) and in the five age  bins that we are considering (red-crossed circles, the bin boundaries are the vertical red lines). In the left-most panel of Figure \ref{nagb}, for each cluster we adopt the ages listed in Table \ref{data_clusters}. The number of TP-AGB stars per unit luminosity increases from bin age2 to bin age3, again confirming the development of the TP-AGB at these ages. 
It is instructive to see how the interpretation of the same star counts changes when other age indicators are used for the same clusters. This is illustrated in the middle and right-most  panels of Figure \ref{nagb}. 
The clusters' migration from one bin to another impacts on the trend of this ratio with age: for example when using \cite{1995A&A...298...87G} calibration of the $s$ parameter  (middle panel) the bin centered at 0.6 Gyr is populated by many clusters and has a relatively high number of TP-AGB stars per unit luminosity. Instead, using the ages from the CMDs this same bin contains just one cluster with bright AGB stars (namely, NGC 1831) which contains six of them.  If we use the age calibration of the $s$ parameter from \cite{2008MNRAS.385.1535P} (right panel) the bin at 0.6 Gyr still contains NGC 1831, as well as NGC 2134 and NGC 2249, with much less AGB stars per unit L$_V$. This shows how difficult it is to obtain robust information from the available data samples, and emphasizes how important the cluster age determinations are. Thus, the use of age indicators, even though calibrated, can induce spurious results due to the sparse data set.

The number of TP-AGB stars per unit luminosity can be used to derive indications on the lifetime of the TP-AGB phase; inverting Equation 1 \citep{1986ASSL..122..195R}: 

\begin{equation}
t_{AGB}=(N_{AGB}/L_{V}) \times (L_{V}/L_{bol}) \times B(t)^{-1}
\end{equation}
Using Maraston models,   L$_{V}$/L$_{bol}$ $\simeq$0.67 almost insensitive to age;  $B(t)$ increases from 0.8 to 1.9 $\times$ 10$^{-11}$ 
stars/L$_{\odot}$/year as a stellar population ages from 0.05 to 3 Gyr. Thus, a value of N$_{AGB}$/L$_{V}$$=$ 10$^{-4}$ {L$_{V,\odot}$}$^{-1}$, as in our bin at $\sim$ 1 Gyr 
implies an evolutionary lifetime of $\sim$ 4 - 5 Myr.   
As the calibrating data set play the critical role in the way these poorly known stellar evolutionary phases  are implemented in the models, in the next section we discuss the origin of model discrepancies as due to the different calibration procedures adopted by different authors.


%
\subsection{Models discrepancies due to calibration procedures}\label{models}
As introduced in Section \ref{intro}, \cite{1998MNRAS.300..872M, 2005MNRAS.362..799M} models adopted tracks without overshooting [\cite{1992ApJS...78..517C}; \cite{1997A&A...317..108C}; \cite{1997MNRAS.290..515C}; \cite{2000MNRAS.315..679C}] and calibrated the  theoretical TP-AGB fuel consumption using the empirical bolometric contribution by TP-AGB stars as measured in the set of MCs' clusters from 
\cite{1990ApJ...352...96F} averaged in age bins. Empirical C and M-type star spectra were used to describe TP-AGB stars in the models. The color of individual clusters from \cite{1983ApJ...266..105P} and \cite{1968AJ.....73..569V}   were used to assess the extent of color excursion during the AGB phase transition, playing with the C, O spectral sub-types until 
matching the overall trend of individual cluster colors in a color-color diagram 
(see Figure 19 of \citealt{2005MNRAS.362..799M}).

The Padova models, and the EPS based on them, are based on  a synthetic TP-AGB evolution, which results from the integration  of a sequence of stellar envelope models under analytic prescriptions for the growth of the core mass. Many physical processes take place during this evolutionary phase, e.g., the third dredge up, nuclear burning at the bottom of the convective envelope, mass loss, stellar pulsation, etc. Therefore, this kind of models require the specification of several parameters, e.g., the amount of dredge-up at each thermal pulse, which is extremely model dependent, envelope convection controlling the hot bottom burning (HBB) process and the rate of luminosity increase, mass loss; atmospheric opacities with or without grains, several critical nuclear reaction cross sections and other more hidden parameters.
In addition, the behavior of the convective zones, the nuclear burning and the mass loss  conspire to establish the surface chemical composition, which impacts on the surface opacity. In turn, the surface opacity drives the radius and temperature of the modeled star, to which the mass loss rate is extremely sensitive. In turn, the more mass is lost, the more rapid the evolution towards low effective temperatures and  higher  mass loss rates. The evolution of TP-AGB stars is then sensitive to many parameters and  on top of this the comparison of the models to the observations  needs assuming color-temperature transformations, which are very uncertain for such cool stars.  

\cite{2007A&A...469..239M} used a sample of nine LMC and six SMC clusters to constrain the duration of the TP-AGB phase, and the fraction of it spent as a carbon star, in a similar fashion as described at
the end of section \ref{integrated}. The observational data were the number of M-type and C-type stars per unit $L_{V}$ in clusters of different ages. The grid of models by \cite{2007A&A...469..239M} was calibrated to fit these constraints, as well as the luminosity function of the field C stars in the MCs, having assumed a specific star formation history for them. A feature of these models is that they include the effect on surface opacity of the chemical composition variation due to the third dredge up and the HBB process \citep{2002A&A...387..507M}. 
The similarity between the M05 and the \cite{2010ApJ...724.1030G} models is quite puzzling if we consider  that they have been derived with very different procedures, although they are  both ultimately constrained to match  the number counts of bright AGB stars in MCs' clusters.
 
BC03 models include a TP-AGB description using models from \cite{1993ApJ...413..641V} which admittedly failed to reproduce the C star distribution of the MCs. For the spectral library they used theoretical C-type star spectra. As discussed in M05 and \cite{2007ASPC..374..303B}, this recipe leads to a very weak effect from the TP-AGB phase on integrated spectra.

The Flexible Stellar Population Synthesis (FSPS) code \citep{2009ApJ...699..486C} make use of both Padova \citep{2008A&A...482..883M} and BaSTI \citep{2004ApJ...612..168P} stellar evolutionary libraries and the same spectral ingredients as the M05 models. In their calibration with MC globular clusters, \cite{2009ApJ...699..486C} have used the average superclusters by \cite{2008MNRAS.385.1535P} and \cite{2004ApJ...611..270G}  that - as we saw in Figure 3 and 4 - have bluer integrated colors with respect to our averaged colors (which is partly due to the different age binning). 
To match the data, \cite{2009ApJ...699..486C}  modified the TP-AGB phase in the Padova isochrones, by decreasing the bolometric luminosity and increasing the effective temperature of the isochrones. As a result, in their calibrating the AGB  contribution  implies bluer $(V-K)$ colors and occurs at older ages with respect to both the M05 and the  \cite{2010ApJ...724.1030G} models. On the other hand, \cite{2010ApJ...724.1030G} noticed  that very low temperature TP-AGB stars do exist in the MCs, and this modification of the tracks would not account for any of them.
  
\cite{2010ApJ...712..833C} criticized the M05 calibration, showing that her models did not fit the trend of the integrated $(V-K)$ color with age of the same clusters used for her calibration.  
 However, the M05 procedure was not designed to fit the $(V-K)$ color as a function of age. Instead, the progressive reddening of the $(V-K)$ color as the stellar population develops an extended AGB was matched
  on a two color plot, using the $(U-B)$ and $(B-V)$ colors as age indicators. This explains part of the difference between these two calibrations. The main discrepancy, however, is due to the different ages attributed to clusters. M05 used  the SWB-type vs. age relation from the prescriptions of \cite{1990ApJ...352...96F} which, as shown in \ref{indicators}, does not reflect the real age of the clusters. 
With the advent of more recent cluster ages from direct CMD fits, the age of  SWB type IV clusters has been shifted to $\sim 1$ Gyr,  in particular having used tracks with mild overshooting, as adopted in 
\cite{2008MNRAS.385.1535P}  and \cite{2010ApJ...712..833C}.  These differences  account for most of the inconsistency.

To further disentangle the discrepancy in \cite{1998MNRAS.300..872M, 2005MNRAS.362..799M} models claimed by \cite{2010ApJ...712..833C}, in Figure \ref{two_panel} we show the color as a function of age
 for the clusters in \cite{1990ApJ...352...96F} (lower panel), with their SWB ages, as this is the set of data used by 
 \cite{1998MNRAS.300..872M, 2005MNRAS.362..799M} to calibrate the TP-AGB fuel, and the age vs color for the clusters in our Table \ref{data_clusters} (upper panel).
 From this figure it is clear that \cite{1998MNRAS.300..872M, 2005MNRAS.362..799M} models are too red from their own calibration. However, the misalignment between the models and the data is larger
 when considering the new accurate CMD age determinations (upper panel) which were not available at the time  \cite{1998MNRAS.300..872M, 2005MNRAS.362..799M} models came out.   
 We also note that, given the rather sparse and low-quality age data available at the time, the M05 models were not calibrated using the (V-K) vs age relation but instead using the (V-K) vs (B-V) two-color plot as seen in Figure 20 of M05. 
 
 Assuming the new set of calibrating data we present in this paper, we now briefly explore how the \cite{2005MNRAS.362..799M} models could be modified to better match the average trend implied by the new calibration.
The new models from Maraston (M13, in preparation) are depicted in Figure~\ref{color_models} as a solid golden line. 

In calculating these new set of models, the fuel consumption during the TP-AGB phase has been set to zero up to an age of 0.6 Gyr. 
This requirement is due to the different age calibration shown in this paper, which pushes some clusters towards older ages, as we extensively discussed. 
This choice is mainly justified by cluster NGC 1831 that, with an age of $\sim$0.7 Gyr, presents about six AGB stars \citep{2006ApJ...646..939M} while younger clusters in bin age2 present either non or an insignificant number of AGBs \citep{1990ApJ...352...96F}.
At ages $\lesssim$0.6 Gyr the observed colors from the new data/new age calibration are just consistent with the sole contribution by the RGB and Early-AGB in the models, so that the TP-AGB phase should provide a negligible contribution. Whether this age onset - based on the new age calibration of the MC clusters - corresponds to the most appropriate one in nature remains to be decided. 
This point will be expanded upon in M13.

At older ages, the TP-AGB fuel consumption  has been somewhat reduced with respect to the original calibration by Maraston (1998), within the error bars of that calibration, and the choice of empirical TP-AGB stars used to construct the models has also been modified. A detailed description of the new models and their effect on the interpretation of galaxies will be presented  in the forthcoming dedicated article (M13).

Comparing M13 and FSPS models, it is noticeable that the adopted datasets contributed to the difference between M05 and FSPS models as M13 are now close to the FSPS ones  (at least for ages older than $\sim$1 Gyr).

\section{Discussion and Conclusions}\label{discussion}

Evolutionary Population Synthesis models provide us with the key tool for our attempts to measure stellar masses, star formation rates, ages, and metallicities of galaxies through cosmic time. Yet they need to be empirically calibrated, which is typically accomplished using nearby star clusters, assumed to be `simple stellar populations'. The star clusters to which we have access are far from providing an ideal database for such calibrations: we lack important age-metallicity combinations; the clusters form an inhomogeneous sample; and they are limited by stochastic fluctuations in energetically important but short-lived evolutionary phases. All of this piles up uncertainties in the model calibrations. To address these problems, and to provide a more robust calibration for intermediate-age stellar populations, in this paper we cull a sample of 43 Magellanic Cloud (MC) stars clusters in the age range $\sim 0.05-3$\,Gyr, using the very latest age determinations. For some clusters, their total infrared luminosity can be dominated by the presence of just a few TP-AGB stars, leading to large stochastic color  variations among clusters of similar age. We have alleviated this problem by stacking the clusters into five distinct age bins, leading to a reasonably well-sampled TP-AGB contribution for each bin. We ensured that our results are robust with respect to our choice of binning. (Note that the age spread of up to $\sim 300$\,Myr reported for some MC clusters older than $\sim 1$\,Gyr are smaller than the width of our chosen bins.)


From the analysis of the luminosity of the clusters, both individually and stacked, we found a clear difference in the behavior of the near-IR and the optical light, reflecting the different sensitivity of the $V$ and the IR bands to the presence of bright AGB stars. The behavior of the clusters' color as a function of age -- key for calibrating EPS models -- shows that clusters become abruptly red in all of the studied colors [$(V-K)$, $(J-K)$ and $(V-I)$] over the age range $\sim 0.6$ to $\sim 1$\,Gyr. This sudden reddening of the colors is  attributed to the appearance of a well developed TP-AGB phase. Clusters older than $\sim$1\,Gyr present a nearly constant red color, within the errors.  

When comparing the color data with the different EPS models from M05, FSPS, BC03 and Girardi et al., we find that no model gives a perfect fit to the data. For ages older than $\sim 1$\,Gyr, M05 and Girardi et al. models  run through the upper end of the error bars whereas BC03 and FSPS models run through their lower end. However, for younger ages both the M05 and Girardi et al models lie well above the upper boundary of our data, being too red; while the BC03 and FSPS models agree with the cluster colors.  
It is important to note that  \cite{2012arXiv1212.5381B} showed that the luminosity functions computed with the BC03 models are too blue compared with the observations.

We proved that the discrepancies between M05 and FSPS models discussed in \cite{2010ApJ...712..833C} are largely due to the cluster ages used by M05 for the model calibration, i.e., the cluster ages from \cite{1990ApJ...352...96F} that were available at that time. Subsequent and more accurate age determinations based on isochrone fits to CMDs indicate that several of the clusters from \cite{1990ApJ...352...96F} are now systematically older, by up to a factor of $\sim 3$ than previously found.

The findings that the TP-AGB phase appears to develop at ages between $\sim 0.6$ and $\sim 1$\,Gyr may have important  implications for stellar evolution. In clusters of $\sim$\,0.3 Gyr, the turn-off mass is $\sim 3$\,M$_\odot$ and this is widely considered the lower mass limit for the HBB process to operate (\citealt{1981A&A....94..175R}; \citealt{2009A&A...499..835V}), thus leading to a prompt abortion of the TP-AGB phase.
Pushing instead the development of the TP-AGB to substantially older ages would imply that the HBB process, if indirectly responsible for the termination of the TP-AGB phase, would need to start operating on stars of initial mass above $\sim 2-2.5$\,M$_{\odot}$. Therefore, it should be explored whether AGB stellar models can be made compliant with this constraint.
This potentially has several important ramifications, ranging from a sizable reduction in the production of primary nitrogen by these kind of stars over previous estimates (e.g.  
\citealt{1981A&A....94..175R}; \citealt{2010MNRAS.403.1413K}), to a potentially important effect on the chemical composition of the AGB ejecta (out of which the multiple populations of Galactic globular clusters may have formed; e.g. \citealt{2008MNRAS.391..354R}; \citealt{2008A&A...479..805V}). It is beyond the aims of the present paper to explore these implications in any detail but it is important to call attention to these aspects of more general interest for stellar evolution and the chemical evolution of stellar clusters and galaxies.	
 
Finally, it is important to stress that the existing photometric data for MC clusters and the ages derived from them are still far from being fully homogeneous in quality and accuracy. A dedicated effort to establish such a homogeneous database would be of great value for the calibration of EPS models and and for a better understanding of advanced evolutionary stages.

\acknowledgments

We thank the referee, Charlie Conroy, for his useful and thorough comments. C.M.C. and N.E.D.N. thank the kind hospitality during their visit at the University of Portsmouth.
N.E.D.N. thanks the warm reception of the Osservatorio Astronomico di Padova. C.M. acknowledges support from the STFC Consolidated Grant.
 


\bibliographystyle{apj}
\bibliography{refs}{}
 
\newpage

\begin{deluxetable*}{cccccccccccccc}
\tabletypesize{\scriptsize}
\tablecaption{Data from the MCs' star clusters.\label{data_clusters}}
\tablewidth{0pt}
\tablehead{
\colhead{Cluster} & \colhead{V$_{0}$} &\colhead{(V-R)$_{0}$}&\colhead{(V-I)$_{0}$} &\colhead{(V-J)$_{0}$} &\colhead{(V-H)$_{0}$} &\colhead{(V-K)$_{0}$}&\colhead{A$_{V}$}&\colhead{Aperture}&\colhead{Age}&\colhead{JHK}&\colhead{VRI}&\colhead{Age}
\\
&  & & & & & & &[$\arcsec$] &  [Gyr] & reference & reference & reference&}
\startdata
NGC 1644  & 12.43    &  0.316 &   0.632    &   1.234    &   1.622  &    1.718   &     0.39   &      60   &   1.55	 & 1 & 3 & 8\\
NGC 1651  & 12.16    & -- &  0.715    &    2.255  &     2.987 &    3.095   &     0.35   &      100 &         2.00 & 1& 3 & 9\\
NGC 1718  & 11.74    &   --   &  --     &   1.874    &   2.699  &   2.857    &    0.51    &     62    &  2.05 & 1& 4 & 9\\    
NGC 1751  & 11.58    &  0.519  &   1.154    &  2.122     &   2.967  &   3.265    &    0.51    &     60    &   1.40	   & 2 & 3 & 8\\
NGC 1777  &  12.41   &     --   &   --      &   3.460      &   3.888  &   3.994    &    0.39    &     38    &  1.15	      & 1 & 5 & 9\\ 
NGC 1783  &  10.60     & 0.400  & 0.860    &  1.451     &   2.028  &    2.110     &    0.30      &   60      & 1.50   & 1 & 3 &  8\\
NGC 1806  & 11.12    &   0.601   &  1.204     &   2.004    &   2.708  &   2.962    &    0.25    &     60    &   1.50	   &  2& 3 & 8\\
NGC 1831  &  10.70     & 0.206   &  0.442     &   0.954    &   1.432  &   1.598    &    0.39    &     60    &   0.70  &  1 & 3 &  9\\ 
NGC 1846  & 10.81    &  0.449  &  0.979   &   1.809    &   2.541  &   2.832    &    0.45    &     60    & 1.48	    & 2 & 3 & 8\\
NGC 1856  &  9.85     &  --    & --      &   0.932    &   1.298  &   1.435    &    0.22    &     60    &   0.30 & 1  & 6 &  9\\     
NGC 1866  &  9.59   &   0.237 &    0.448   &   0.952    &   1.362  &   1.514    &    0.28    &     60    &   0.20 &1 & 3 &10 \\     
NGC 1868  & 11.14    &  0.240  &   0.556    &  1.058     &  1.496   &  1.622     &   0.39     &    60     &  1.10   & 1 & 3 & 9\\  
NGC 1978  & 9.86    &  0.364  &  0.679   &    1.330     &  1.949   &  2.131     &   0.76     &    60     &    2.00   & 1 & 3 &  8 \\
NGC 1987  & 11.72    &  0.389  &   0.885    &  1.639     &  2.289   &  2.711     &   0.28     &    60     &  1.08   & 1 & 3 &  8\\
NGC 2031  & 10.43    &   --   &  --     & 1.593      &  2.220     &2.315       &  0.40        & 72        &  0.16	    & 1 & 4 & 11\\  
NGC 2058  & 10.34    &  0.154  &    0.370     & 0.882      &  1.380     &1.436       & 0.39       &  60       &  0.14 & 1 & 3 & 12\\    
NGC 2107  & 11.15    &   --   &  --     & 1.202      & 1.783    &  1.990       & 0.36       &  60       &  0.62	 & 1 & 6 &  1\\  
NGC 2108  & 11.82    &   --   &  --     & 1.501      & 2.208    & 2.626      &   0.50       &  62       & 1.03     &  1& 4 &  8\\ 
NGC 2121  & 11.84    &   --   &   --      &  1.589     &  2.203   &  2.699     &   0.53     &    62     &  2.90 & 1 & 4 & 9\\
NGC 2134  & 10.35    &  0.106  &  0.241   &  0.587   &  0.961 &   0.991  &    0.62    &     60    &  0.19 & 1 & 3 & 13 \\   
NGC 2136  &  10.20     & 0.214   & 0.487    &  1.005     &  1.442   &  1.544     &    0.30      &   60      & 0.10 & 2 & 3 & 11\\  
NGC 2154  & 11.71    &  0.391  &   0.857    &  1.679     &  2.417   &  2.853     &   0.39     &    60     & 1.43 &  1& 3 &  8\\
NGC 2155  & 12.27    &  0.488  &  0.783   &   1.407    &   2.031  &   1.944    &    0.43    &     60    &     3.00	& 1 & 3 & 9\\  
NGC 2156  & 11.18    &   --   &  --     &   0.426  &     0.785 &    0.772 & 0.20        & 72        & 0.05 & 1 & 4 & 14 \\	  
NGC 2157  &  9.86     &  --    &    --        &  0.825   &    1.302 &    1.444   &      0.30    &     60    & 0.11	   & 2 & 4 & 15 \\  
NGC 2162  & 12.22    &  0.435  &   0.861    &  1.443     &   2.081  &   2.257    &    0.39    &     60    &  1.25 &  1& 3 &  9\\  
NGC 2164  & 10.01    & --  &  0.215   &   0.384 &  0.672  &  0.733   &      0.30    &      60   &     0.20	  & 2 & 3 & 16\\   
NGC 2173  & 11.92    &  0.458  &   1.014    &  2.266     &  2.884   &    3.050     &   0.39     &   150    &   1.60   & 1 & 3 & 17 \\      
NGC 2190  & 12.55    &   --   &  --     &  1.240       &  1.828   &    2.024   &     0.39   &      61   &    1.10	    & 1  & 5 & 18 \\  
NGC 2193  & 13.03    &   --   &  --     &  1.130       & 1.738    &   1.804    &    0.39    &     38    &   2.20   & 1 & 5 & 19\\
NGC 2203  &  10.90     &  --    & --      & 1.730        & 2.358    &   2.544    &    0.39    &    150   &    1.80 & 1 & 5 & 18\\
NGC 2209  & 12.76    &   --   &  --     &   2.100        & 2.928    &   3.424    &    0.39    &     60    &   1.20	 &  1& 4 &  20\\  
NGC 2213  & 12.08    &  0.496  &  0.960   &    1.790     &  2.597   &   2.872    &     0.40     &    60     &  1.70 & 1 & 3 & 9\\
NGC 2231  & 12.84    &  0.443  &   0.799    &  1.621     &  2.399   &   2.695    &    0.39    &     45    &   1.50 & 1 & 3 &  18 \\
NGC 2249  & 11.84    &   --   &  --     &  0.890       & 1.338    &  1.634     &   0.39     &    72     &    1.00    & 1 & 4 & 9\\   
SL 842       &   13.76 &     --  &   --    &   1.980     &  2.688    &  2.944    &  0.39       &  38       &     2.00 & 1& 5 & 18\\  
Hodge 4     &  12.94     &  --     & --       &  0.950      & 1.228     & 1.544     &  0.39      &  38        &    2.10 & 1 & 5 & 21 \\
Hodge 14   & 13.03   &   --    & --      & 1.090       & 1.728     & 1.604     &  0.39      &  62        &    2.25	  &  1 & 7 & 9\\
NGC 152   &  12.63   &  0.542  &  1.112     & 1.904      &  2.713   &  3.031     &  0.19      &  60       &    1.40   & 1 & 3 &  22\\  
NGC 265   & 11.84    & 0.250   & 0.550    & 1.032      &  2.015   &   2.114    &   0.34     &   60      &    0.32	  & 1 & 3 & 23 \\
NGC 269   &  11.30     & --     & --      & 0.216    & 0.770  & 0.918    &   0.34     &   60      &    0.32	  & 2 & 4 & 24\\   
NGC 411   & 12.02    & 0.518   & 0.823    & 1.579      & 2.211    &   2.430      & 0.17       & 60        &   1.40   & 1 & 3 &  22\\  
NGC 419   & 10.28    & 0.464   & 0.903    & 1.565      & 2.301    &   2.621    &   0.32     &   60      &    1.20	&  2& 3 & 25\\

\enddata
\tablecomments{References: 1: \cite{2006AJ....132..781P}, \cite{2008MNRAS.385.1535P}; 2:  \cite{1983ApJ...266..105P}; 3: \cite{2006MNRAS.369..697G}; 
 4: \cite{1981A&AS...46...79V}; 5: \cite{1996ApJS..102...57B};  6: \cite{1968AJ.....73..569V}; 7: \cite{1975A&A....40..199B};  
8: \cite{2009A&A...497..755M}, CMD fitting, all observed CMDs reach MSTO;  9: \cite {2007A&A...462..139K}, CMD fitting, all observed CMDs reach MSTO; 10: \cite{1999AJ....118.2839T} LFs, CMD reaches MSTO; 
11: \cite{2000A&A...360..133D} isochrone fitting; 12: Sebo, K. 1996, PhD Thesis, isochrone fitting; 13: \cite{1994A&A...284..424V}, isochrone fitting; 14: \cite{2008RMxAA..44...57S}, isochrone fitting, CMD reaches MSTO;
15: \cite{1998AJ....115..592F}, isochrone fitting; 16: \cite{1991A&AS...87..517V}, isochrone fitting; 17: \cite{2003AJ....125..770B}, isochrone fitting, CMD reaches MSTO; 18: \cite{1997AJ....114.1920G}, isochrone fitting; 
19: \cite{2001AJ....122..842R}, isochrone fitting, CMD reaches MSTO; 20: \cite{2012ApJ...761L...5K}, isochrone fitting, CMD reaches MSTO; 21: \cite{2003AJ....125..754W}, CMD fitting, CMD reaches MSTO;
22: \cite{2001AJ....122..220C}, isochrone fitting, CMD reaches MSTO; 23: \cite{2007A&A...466..165C}, isochrone fitting, CMD reaches MSTO; 24: \cite{2006A&A...452..179C}, isochrone fitting, CMD reaches MSTO; 
25: \cite{2008AJ....136.1703G}, isochrone fitting, CMD reaches MSTO.}
\end{deluxetable*}

\begin{deluxetable*}{ccccccccc}
\tabletypesize{\scriptsize}
\tablecaption{Sampled luminosities, masses and expected Number of TP-AGB stars in the MCs' star clusters\label{n_agb}}
\tablewidth{0pt}
\tablehead{
\colhead{Interval}& range of ages & Number of AGBs & K-band & V-band &  Mass &
\\
 name & [Gyr] & assuming t$_{AGB}$=1 Myr & Luminosity & Luminosity & from $\frac{M}{L_{\rm V}}$}
\startdata
age1 & 0.05 $\leq$ age $<$ 0.3 &   --- &  1.292E+06& 1.392E+06 &       0.3E+06   \\
age2 & 0.3 $\leq$ age $\leq$ 0.9  & 14.0 & 5.818E+05 & 5.776E+05 &  0.28E+06   \\
age3 & 0.9 $<$ age $\leq$ 1.4 &  15.5 & 1.809E+06& 6.456E+05 &   0.3E+06    \\
age4 & 1.4 $<$ age  $<$ 2 &  13.0 & 1.484E+06 & 5.422E+05 &     0.54E+06      \\
age5 & 2 $\leq$ age $\leq 3$&   10.2  & 8.817E+05 & 4.275E+05 &       0.62E+06     \\
\enddata
\end{deluxetable*}


\end{document}